# Health App Reviews for Privacy & Trust (HARPT): A Corpus for Analyzing Patient Privacy Concerns, Trust in Providers and Trust in Applications

Timoteo Kelly[1],BS, MSHI; Abdulkadir Korkmaz[2]; Samuel Mallet[2], BS; Connor Souders[2]; Sadra Aliakbarpour[3]; Praveen Rao[1,2], BS, MS, PhD

[1]Institute for Data Science & Informatics, University of Missouri, Columbia, MO, USA, Email: tkb5b@missouri.edu

[2]Dept. Electrical Engineering & Computer Science, University of Missouri, Columbia, MO, USA, Email: ak69t@missouri.edu, Email: sm9dm@missouri.edu, Email: cosbdf@missouri.edu, Email: praveen.rao@missouri.edu

[3]Dept. of Computer Science, Yale University, New Haven, CT, USA, Email: sadraprivate@gmail.com

**Corresponding Author:**

Timoteo Kelly
Email: tkb5b@missouri.edu

## Abstract

**Background:** User reviews of Telehealth and Patient Portal mobile applications (apps) hereon referred to as electronic health (eHealth) apps are a rich source of unsolicited patient feedback, revealing critical insights into patient perceptions. However, the lack of large-scale, annotated datasets specific to privacy and trust has limited the ability of researchers to systematically analyze these concerns using natural language processing (NLP) techniques.

**Objective:** This study aims to develop and benchmark *H*ealth *A*pp *R*eviews for *P*rivacy & *T*rust (*HARPT*), a large-scale annotated corpus of patient reviews from eHealth apps to advance research in patient privacy and trust.

**Methods:** We employed a multistage data construction strategy. This integrated keyword-based filtering, iterative manual labeling with review, targeted data augmentation, and weak supervision using transformer-based classifiers. A curated subset of 7,000 reviews was manually annotated to support machine learning model development and evaluation. The resulting dataset was used to benchmark a broad range of models.

**Results:** The HARPT corpus comprises 480,000 patient reviews annotated across seven categories capturing critical aspects of trust in the application (TA), trust in the provider (TP), and privacy concerns (PC). We provide comprehensive benchmark performance for a range of machine learning models on the manually annotated subset, establishing a baseline for future research.

**Conclusions:** The HARPT corpus is a significant resource for advancing the study of privacy and trust in the eHealth domain. By providing a large-scale, annotated dataset and initial benchmarks, this work supports reproducible research in usable privacy and trust within health informatics. HARPT is released under an open resource license.

# Introduction

Electronic health (eHealth) applications (apps) play a growing role in healthcare delivery, allowing patients to schedule appointments, view medical records, participate in telemedicine, and manage chronic diseases directly from their smartphones. As these platforms collect and process increasingly sensitive personal health information, questions surrounding privacy, trust, and data governance have become central to patient adoption and ongoing use [1] [2].

We refer to our dataset as consisting of *eHealth* apps, a category that includes apps offering telehealth services or patient portal functionalities. Although these functions can exist separately or combined within a single platform, they are commonly grouped under the broader concept of eHealth which includes digital tools that support remote care and access to personal health data.

Despite the ubiquity of these eHealth apps, there is a lack of publicly available large-scale datasets that capture user perspectives on privacy and trust. Existing research relies heavily on surveys, interviews, or proprietary datasets, limiting reproducibility and long-term reuse [1–3]. This presents a critical gap: without open, reusable resources, it is difficult to benchmark methods, compare findings, or preserve evidence of evolving user concerns in health technologies.

To address this need, we introduce *H*ealth *A*pp *R*eviews for *P*rivacy and *T*rust (*HARPT*), a corpus of over 480,000 app store reviews from patient portal and telehealth platforms. These reviews capture every- day frustrations and concerns, such as “Can’t sign in to my patient portal” or “App very slow, doctors hung up.” These examples illustrate everyday issues of reliability, data quality, and access that shape user trust and privacy concerns (PC), which HARPT systematically captures. To our knowledge, HARPT is the first resource to jointly encode PC alongside two dimensions of trust: trust in providers (TP) and trust in application (TA). The dataset was developed through a multistage pipeline combining keyword filtering, multi-rater annotation, targeted data augmentation, and weak supervision with transformer models. In addition to releasing the dataset, we provide baseline experiments across traditional classifiers and transformer-based models. By making both the corpus and benchmarks openly available, HARPT supports comparative evaluation, contributes to the ethical preservation and reuse of user-generated data, and enables reproducible research in usable privacy, natural language processing (NLP) and health informatics.

## Related Work

Privacy and security in eHealth and digital health have been studied through conceptual frameworks [4, 5], trust quantification models [6] , and reviews of user attitudes toward data sharing [7] or privacy in developing countries [8] . While valuable, these studies rely mainly on surveys or theoretical methods and rarely release open datasets.

A smaller body of work has analyzed user reviews. Large-scale studies have examined app reviews for security and PC [9–11] , while other approaches include unsupervised summarization [12], ethical concern mining [13], semi-supervised classification [14], misinformation detection [15], and COVID contact-tracing analysis [16] . However, most use proprietary or non-health datasets, focus on isolated aspects of trust or privacy, and do not release reproducible resources.

HARPT addresses these limitations by providing a large-scale, openly licensed dataset labeled for both privacy and dual trust constructs. By aligning with the FAIR principles and releasing models alongside data, HARPT advances reproducible, long-term research.

# Methods

We collected 457,165 user reviews from the Google Play store, including both patient portal and telehealth services. A keyword-based filter retained reviews mentioning privacy or trust concepts for annotation. From this pool, 4,000 reviews were randomly sampled and manually labeled into seven mutually exclusive categories: competence, reliability, support, risk, ethicality, data quality, and data control. Labels were assigned using a three-pass protocol with majority voting across annotators, yielding high reliability (Fleiss’ Kappa=0.877) [17].

## Theoretical Framework for Privacy and Trust

Our taxonomy draws on established frameworks in privacy and trust. For *PC,* we reference IUIPC [18], ubiquitous computing concerns [19], and healthcare-specific trust models [20–22], which motivate the *data control* and *data quality* categories. For *TP,* we build on Mayer et al. [23], who identify ability, benevolence, and integrity as key antecedents of trust, and Hosmer et al. [24] , who emphasizes fairness and ethical responsibility.

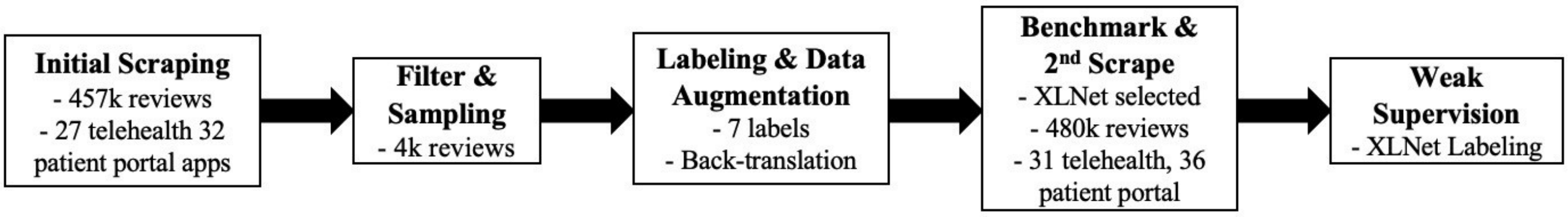

**Figure 1.** Overview of the HARPT construction pipeline.

| Label | Description + Example |
|---|---|
| **Data Control** | Concerns about consent, third-party access, or control over personal data. "It's great the app lets me review and delete my medical records whenever I want." |
| **Data Quality** | Issues with incorrect, outdated, or incomplete health data. "The app is showing someone else's health record!" |
| **Reliability** | App crashes, login failures, or inconsistent functionality. "I can't login! Every time I enter my username and password it just loads and loads." |
| **Support** | User experience with help services or communication. "Customer support responded to my query within a few hours and helped me resolve my issue." |
| **Risk** | Worries about security, breaches, or misuse of data. "I'm worried about how the app stores my personal health records." |
| **Competence** | Perceptions of provider professionalism and care quality. "I received tailored advice based on my health data and it helped me manage my condition better." |
| **Ethicality** | Judgments of provider fairness, transparency, or responsibility. "I love that the app asks for my consent before collecting my data." |

**Table 1.** HARPT labels with description and examples.

These perspectives map directly to our labels: *competence* and *ethicality*. For *TA*, we draw on McKnight et al.'s work on trust in technology [25], which emphasizes beliefs about a system's reliability, helpfulness, and functionality, and Gefen et al.'s integration of trust with technology acceptance constructs [26], which highlight how risk, perceived usefulness, and trust in the technology user behavior. These theories inform our labels of *reliability*, *support*, and *risk*.

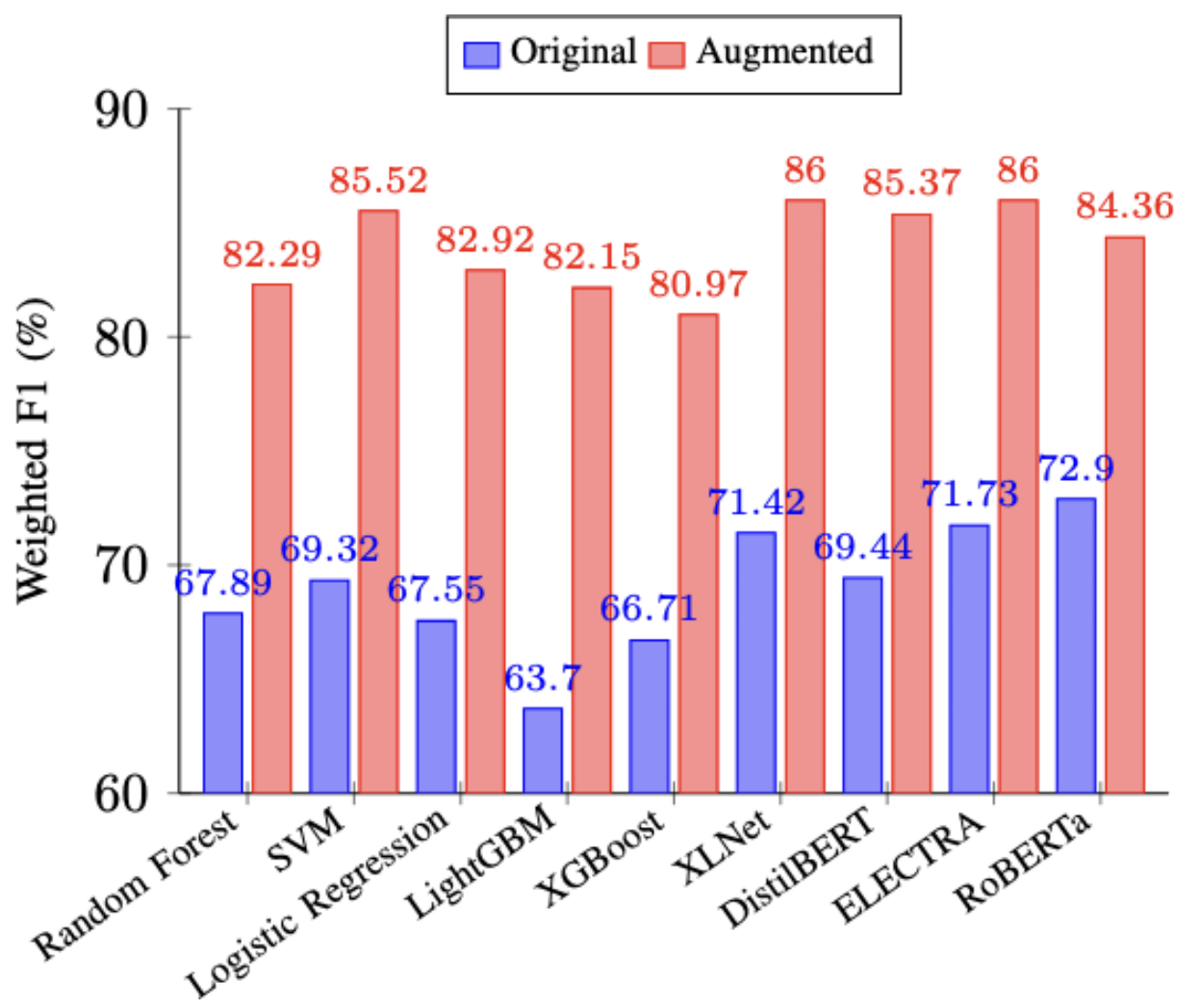

**Figure 2.** Model performance before and after back-translation augmentation.

### Data Augmentation and Evaluation

The ground truth set was imbalanced, with *competence* dominating. We applied back-translation via French, Spanish, and German to oversample minority classes and downsampled majority ones, yielding a balanced set of 7,000 reviews. For example, the review "They are awesome and always fast in responding" was back-translated via Spanish as "They are great and reply quickly," introducing lexical variation while preserving the original meaning. *BLEU score* yielded a similarity of *30.43*, indicating semantic fidelity. We benchmarked nine models – Random Forest [27], SVM [28], Logistic Regression [29], LightGBM [30], XGBoost [31], XLNet [32], DistilBERT [33], ELECTRA [34], and RoBERTa [35] . Results (see Figure 2) show transformers outperform classical models, with XLNet and ELECTRA achieving the best F1-scores, guiding our choice of XLNet for weak supervision.

Using XLNet, we weakly labeled 480,450 reviews collected from 67 mobile health apps between 2011 and 2025. Reviews are concise (median length 10 words, mean 16.5, max 675) and star ratings are skewed toward positive sentiment (65.4% five-star, 16.7% one-star). Trust and privacy categories are unevenly distributed: competence (44.3%) and reliability (27.5%) dominate, while minority classes such as risk (1.4%) and ethicality (1.3%) are rare. Confidence scores ranged from 0.20–0.999 (median=0.96), with 51.9% $\geq$0.9, 20.9% between 0.7–0.9, and 27.1% $<$0.7. Per-class results (Figure 3) show high F1 for minority categories but more difficulty with competence.

Coverage spans both patient portal (64.8%) and telehealth platforms (35.2%), with reviews concentrated in widely used apps such as *MyChart* (19.4%), *FollowMyHealth* (14.7%), *healow* (9.8%), *Practo* (9.1%), and *Doc- tor on Demand* (7.3%), which together account for over 60% of the corpus.

This scale and diversity support *longitudinal analyses* of trust and privacy over 14 years, *qualitative studies* of user perceptions, *benchmarking* for privacy-aware text analy-sis, and *evaluation of models* for sensitive domains. The class imbalance further makes HARPT a valuable testbed for methods addressing skewed data.

Beyond quantitative evaluation, HARPT captures the richness of user experiences in eHealth contexts. Table I illustrate how reviews naturally surface distinct trust and PC. These examples highlight how reviews embed concrete user concerns, ranging from functional reliability to deeper issues of fairness, professionalism, and data stewardship.

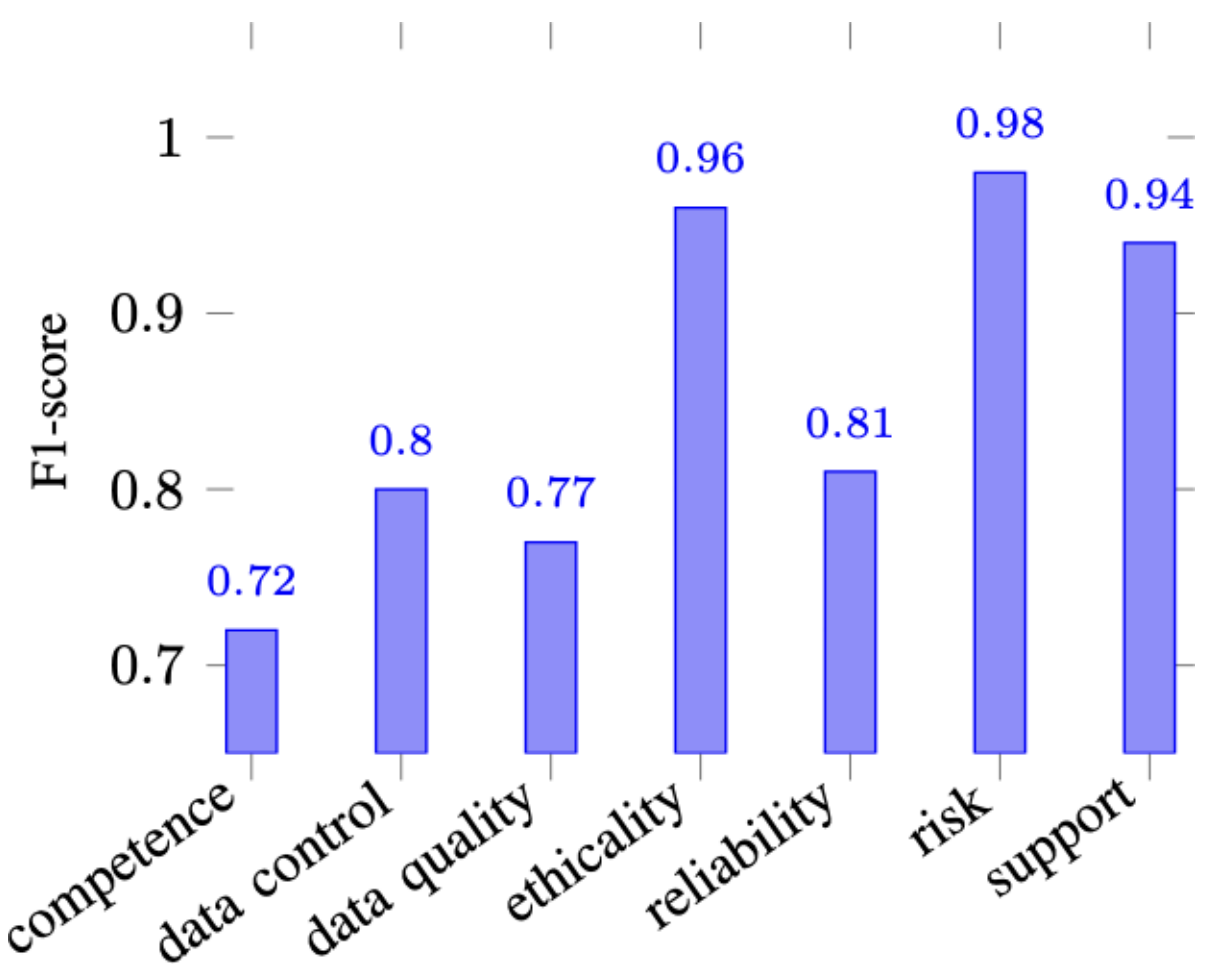

**Figure 3.** Per-class F1-scores on the large-scale dataset (weakly labeled by XLNet).

HARPT, including the weakly labeled dataset and ground truth subset, is publicly available via Harvard Dataverse [36]. The fine-tuned XLNet is released through Hugging Face [37] , under CC-BY 4.0.

## Results

To provide baseline performance on the HARPT dataset, we trained and evaluated both classical machine learning models and state-of-the-art transformer-based models.

For the classical models, we implemented a RF classifier using Term Frequency-Inverse Document Frequency vectorized features extracted from the text reviews. Hyperparameters including number of estimators, maximum depth, and minimum samples per split were tuned via grid search.

For the transformer-based models, we fine-tuned both DistilBERT and RoBERTa on the annotated training data. Fine-tuning was performed using mixed precision training with GPUs, utilizing early stopping and learning rate schedulers to optimize convergence. All hyperparameters were tuned using the sweeps feature on the Weights and Biases platform [38].

| Model | Accuracy | F1 | Precision | Recall |
|---|---|---|---|---|
| Random Forest | 94.00% | 93.96% | 94.05% | 94.00% |
| DistilBERT | 91.25% | 91.27% | 91.32% | 91.25% |
| RoBERTa | 89.02% | 89.04% | 89.13% | 89.02% |

**Table 2.** Benchmark performance on HARPT dataset.

These results (see Table II) demonstrate that both classical and transformer-based models achieve strong performance. Although RF achieves the highest overall F1, transformer models offer greater flexibility for transfer learning and domain adaptation, making them valuable baselines for future research.

To better understand model behavior, we inspected common misclassifications on the ground truth set.

We observed that *reliability* and *support* were frequently confused, as both often reference user frustrations with login failures or lack of assistance.

Similarly, *competence* reviews occasionally overlapped with *data quality*, where patients conflate provider expertise with accuracy of information presented in the app.

Such overlaps reflect the blurred boundaries between user experiences of technical reliability, provider professionalism, and informational correctness in real-world contexts.

These findings highlight that model errors are not purely technical but also reveal conceptual ambiguities in how users articulate trust and PC. This underscores HARPT's value not only as a benchmark for model performance but also as a lens into how trust and privacy are expressed in real-world contexts.

We also tested the state-of-the-art generative transformer GPT-4 [39], which at the time represented the most advanced LLM available. When prompted directly, however, it performed poorly (F1 = 0.44, accuracy = 0.44). This outcome likely reflects the domain-specific and noisy nature of the data as well as the need for task-specific fine-tuning. Given its cost, limited reproducibility, and weaker performance, we focused our analysis on open-source classifiers. Future work may revisit LLMs for tasks where their generative capabilities may offer complementary value.

All data were collected from public Google Play reviews. Usernames were anonymized. We verified that no protected health information (PHI) was retained. The dataset is released solely for research use under Creative Commons 4.0.

## Discussion

HARPT not only benchmarks model performance but also opens new avenues for cross-disciplinary research on trust- and privacy-aware NLP in health informatics and digital libraries. While it provides the largest publicly available corpus of eHealth app reviews labeled for trust and privacy, several limitations should be noted. The dataset is restricted to patient portal and telehealth apps in the Google Play Store. As such, findings may not generalize to other categories of eHealth tools or to reviews from other marketplaces such as Apple's App Store. In addition, the weak supervision pipeline, though carefully validated, unavoidably introduces some noise into the large-scale labels. Confidence scores are provided to mitigate this limitation, but users of the dataset should be mindful of uncertainty when applying labels to downstream tasks.

The dataset is imbalanced, with *competence* and *reliability* dominating. While this reflects real-world prevalence of user concerns, it poses challenges for modeling and may require reweighting or resampling. To address this in the ground truth subset, we applied back-translation to augment minority classes; while BLEU scores confirmed preservation of meaning, some semantic drift may remain. Reviews are short and linguistically variable, which can challenge fine-grained interpretation, and the dataset reflects self-selected users who may not fully represent the broader patient population.

HARPT is the first large-scale, openly available corpus of eHealth reviews labeled for privacy and trust, built through manual annotation, data augmentation, weak supervision, and large-scale scraping. Along with the weakly labeled corpus, we release a ground truth subset and benchmark results from multiple classifiers, providing strong baselines for reproducible research in digital libraries, health informatics, and NLP. Future work will expand to additional marketplaces and update the dataset to capture evolving trends in trust and privacy.

## Acknowledgements

The first author, Timoteo Kelly, acknowledges support of the National Science Foundation and U.S. Office of Personnel Management CyberCorps Grant No. #1946619. The second author, Abdulkadir Korkmaz, acknowledges support from the Republic of Türkiye.